\documentclass[]{spie}  

\newcommand{\um}{$\mu$m}

\usepackage{amsmath,amsfonts,amssymb}
\usepackage{graphicx}
\usepackage[table, x11names]{xcolor}
\usepackage[colorlinks=true, allcolors=blue]{hyperref}
\usepackage[rightcaption]{sidecap}

\title{Update on the Magellan InfraRed Multi-Object Spectrograph (MIRMOS)}

\author[a]{Maren Cosens}
\author[a]{Nicholas P. Konidaris II}
\author[a]{Gwen C. Rudie}
\author[a]{Andrew B. Newman}
\author[b]{Leon Aslan}
\author[c]{Robert Barkhouser}
\author[a]{Christoph Birk}
\author[d]{Julia Brady}
\author[e]{Mathias Gumy}
\author[a]{Tyson Hare}
\author[f]{Stephen C. Hope}
\author[a]{Charlie Hull}
\author[a]{Karim Kaismoune}
\author[a]{Daniel D. Kelson}
\author[a]{Gerrad Killion}
\author[g]{Alicia Lanz}
\author[h]{Jocob McCloskey}
\author[a]{Solange V. Ramirez}
\author[e]{Cyril Rupf}
\author[a]{William Schoenell}
\author[i]{Patricio Schurter}
\author[f]{Stephen A. Smee}
\author[e]{Peter Spanoudakis}
\author[a]{Jason E. Williams}

\affil[a]{Carnegie Science, The Observatories, 813 Santa Barbara Street, Pasadena, CA 91101, USA}
\affil[b]{Turion Space, Irvine, CA, USA}
\affil[c]{LCS Optics LLC, Parkton, MD 21120, USA}
\affil[d]{The Ohio State University, Columbus, OH 43210, USA}
\affil[e]{CSEM SA (Centre Suisse d’Electronique et de Microtechnique), Jaquet-Droz 1, CH-2000 Neuchâtel, Switzerland}
\affil[f]{Johns Hopkins University, Department of Physics and Astronomy, 3701 San Martin Drive,
Baltimore, MD 21218, USA}
\affil[g]{Capella Space, 438 Shotwell St., San Francisco, CA, 94110}
\affil[h]{Penn State University, University Park, PA 16802, USA}
\affil[i]{Giant Magellan Telescope Organization, 300 N Lake Ave, 14th Floor, Pasadena, CA 91101}

\authorinfo{Further author information: (Send correspondence to M.C.)\\M.C E-mail: mcosens@carnegiescience.edu}

\begin{document} 
\maketitle

\begin{abstract}
The Magellan InfraRed Multi-Object Spectrograph (MIRMOS) will be a next generation multi-object (MOS) and integral field spectrograph (IFS) for the 6.5m Magellan telescopes at Las Campanas Observatory in Chile. MIRMOS will perform $\rm R\sim3700$ spectroscopy over a simultaneous wavelength range of $\rm 0.886 - 2.404\mu m$ (Y, J, H, and K bands) in addition to imaging over the range of $\rm 0.7 -0.886\mu m$. Target selection in the MOS mode is achieved through a cryogenic mechanism capable of making up to 92 slits over a $\rm13'\times3'$ field of view (FoV). This mechanism can be reconfigured in real time to adjust slit widths for seeing conditions, change to a different mask, or form a long slit. The IFS mode of operation for MIRMOS will be achieved via an image slicer style integral field unit (IFU) which will provide an $\rm\sim18''\times26''$ FoV made up of $\rm0.84''\times26''$ slices. MIRMOS’s design will allow it to address a wide range of science cases from performing spectroscopy of high-redshift galaxies as well as high signal-to-noise transmission spectroscopy of exoplanet atmospheres with the MOS/long slit mode, while also having the largest FoV IFS operating on a large telescope at these wavelengths. We will describe here the design of the MIRMOS instrument -- now at the end of the preliminary design phase.
\end{abstract}


\section{INTRODUCTION} \label{sec:intro}
MIRMOS is being designed as a powerful new facility instrument for the 6.5m Magellan telescopes. As such it is designed to have capabilities which will allow for advances in a wide range of science cases from studying distant galaxies to the atmospheres of exoplanets.

MIRMOS will provide improvements over existing near-infrared (near-IR) spectrographs in three key areas. It will perform spectroscopy over the full Y, J, H, and K bands simultaneously through use of a dichroic tree, improving cross band calibration and survey efficiency. A selectable engineered diffuser and custom detector readout will enable high spectrophotometric precision for performing exoatmosphere transmission spectroscopy from the ground. And lastly, a seeing limited slicer integral field unit will provide an unprecedented field of view at near-IR wavelengths.

MIRMOS is currently nearing the end of the preliminary design phase and has been described in previous SPIE proceedings\cite{Konidaris2020, Konidaris2022, Konidaris2024}. Here we will focus on the updates and progress on the development since 2024. The key science cases driving the instrument requirements will be discussed in Section \ref{sec:science} with Section \ref{sec:design} providing an overview of the system design and detailed discussion of key components which have seen significant progress since the last update.

\section{KEY SCIENCE CASES \& REQUIREMENTS} \label{sec:science}
While MIRMOS is being designed as a facility instrument, the scientific requirements are driven by three main areas of research: surveys of high redshift galaxies (MOS), study of the extended CGM conditions (IFS), and measurement of the molecular content of exoplanet atmospheres (spectrophotometry). These science cases drive the requirements of the instrument and inform the design choices in MIRMOS. Table \ref{tab:reqs} highlights the key instrument requirements and the science case(s) driving them.

\begin{table}[h]
\centering
\caption{Key instrument parameters and science requirements flowdown} 
\label{tab:reqs} 
\includegraphics[width=\linewidth]{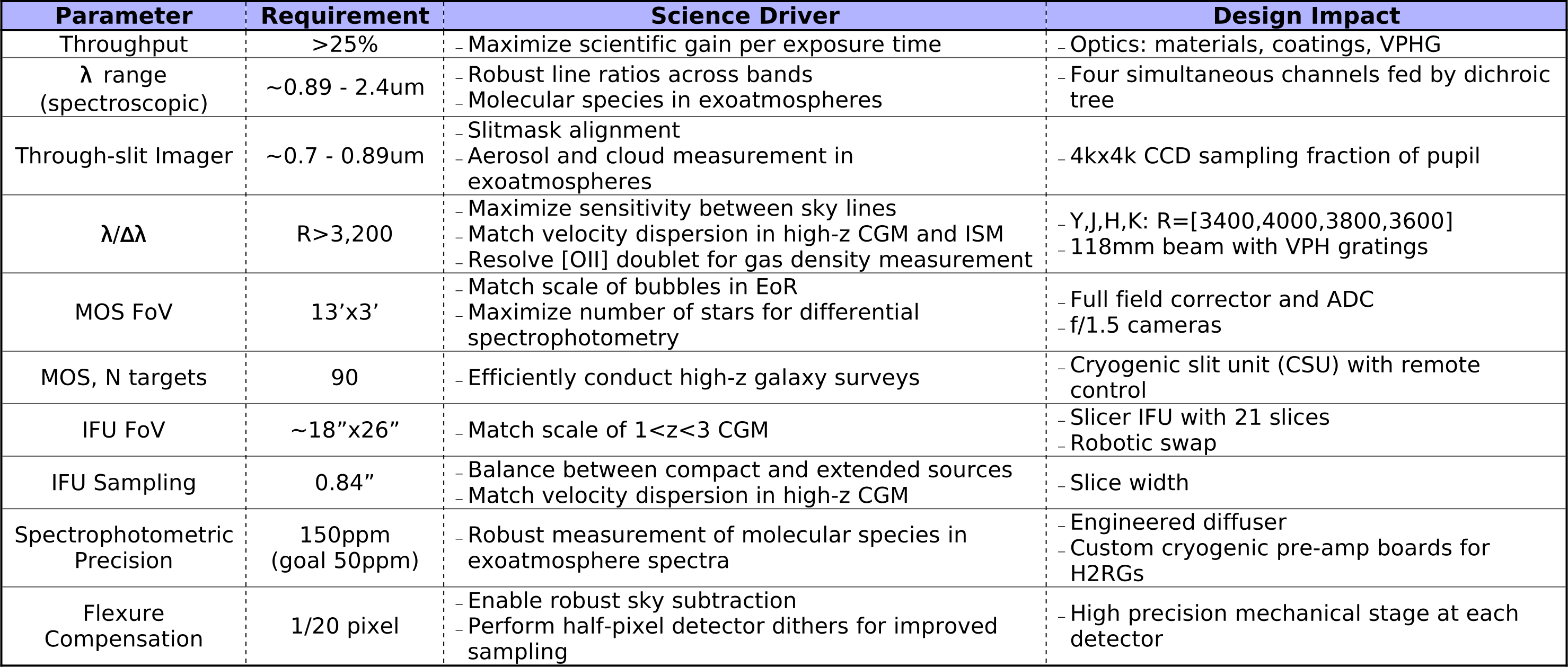}
\end{table}

The most unique and ambitious aspects of MIRMOS are its FoV (particularly the IFU), simultaneous wavelength coverage, and spectrophotometric precision. The scientific need informing these particular requirements is highlighted here.

To study the conditions of star forming galaxies at cosmic noon, observations of emission line ratios which span multiple bands are required. This presents a complication with current instruments in which these bands cannot be observed simultaneously. Variations of the sky background, weather, and seeing conditions with time limit the accuracy of these diagnostic line ratios that are used to probe key galaxy properties like star-formation rate, chemical abundances, and the stellar population. Simultaneous coverage of the full near-IR bandpass is required to improve the accuracy of cross-band calibration while providing an increase in observational efficiency for large MOS surveys.

Wide-field optical IFS's\cite{Morrissey2018, Bacon2010, Furesz2020, Mateo2022, Hill2018} have been revolutionary for the study of the circumgalactic medium (CGM) and interstellar medium (ISM) in nearby galaxies. However, near-IR IFS's have generally been paired with adaptive optics systems and have therefore prioritized finer sampling over a wide FoV. For observation of extended objects, these current near-IR instruments are simply not practical, but there are key shortcomings to observations being limited to optical wavelengths. Observations of the CGM at cosmic noon -- the peak of structure formation -- are primarily limited to Ly$\alpha$ and other resonant transitions in the rest-UV, meaning that the distribution and conditions of this gas is largely unconstrained. Wide-field IFS capability in the near-IR would measure rest-optical emission lines and allow the first robust probes of the morphology, kinematics, and ionization conditions in the CGM above $\rm z\sim1$.

Most studies of exoplanet atmospheres are conducted from space, where there is no impact of our own atmosphere on the measured signal. However, the amount of observing time available from space based facilities like JWST is limited; particularly when it comes to re-visiting previously observed planets to study exoplanet atmospheric variability, or weather. A ground-based instrument on the other hand, would be poised to study a large sample of exoplanet atmospheres to study population trends and time-variable phenomena if similar precisions can be achieved. MIRMOS will seek to do this in two ways: first, by performing differential spectrophotometry with simultaneous reference star observations, and second, by using an engineered diffuser which produces a larger and more stable point spread function (PSF).

\section{INSTRUMENT DESIGN}\label{sec:design}
\begin{figure}[h]
    \centering
    \gridline{\fig{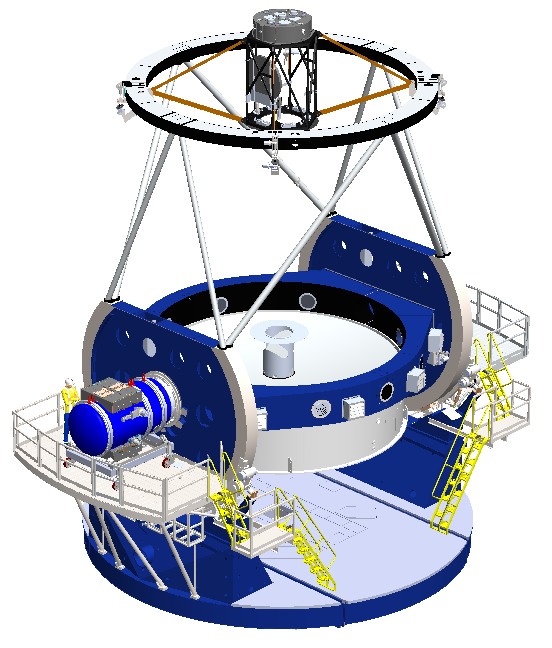}{0.5\textwidth}{}
    \fig{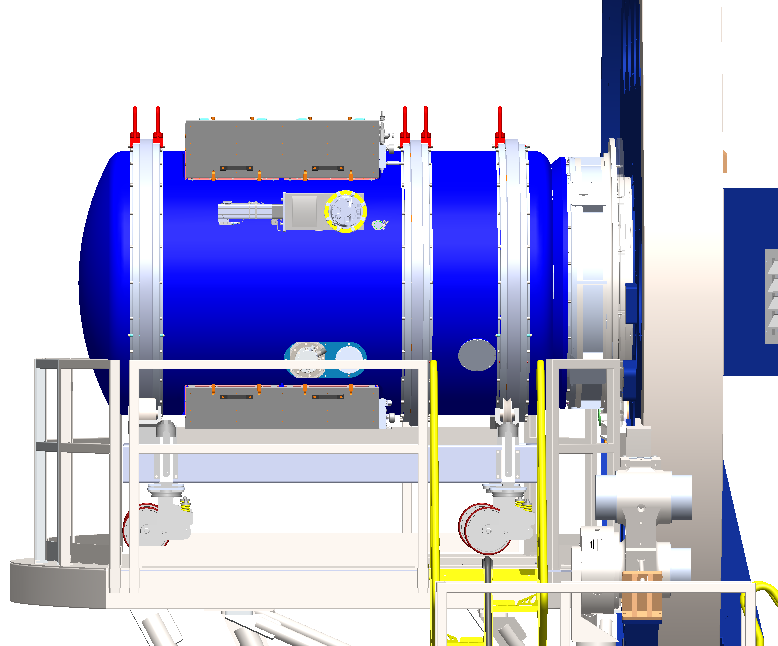}{0.4\textwidth}{}}
    \vspace{-1.5em}
    \caption{CAD model of MIRMOS mounted on the Nasmyth platform of Magellan. The zoom-in view on the right shows MIRMOS mounted to the instrument rotator built into the Magellan elevation disk. An astatic support at the back of the MIRMOS cart provides support for the weight of the instrument. The red hooks at the top of the instrument are for installation and will be removed during operation.}
    \label{fig:Mag_MIRMOS}
\end{figure}

MIRMOS is a wide-field multi-object and integral field spectrograph covering the full near-IR bandpass. It will operate at the Nasmyth platform of the 6.5m Magellan telescopes as shown in Figure \ref{fig:Mag_MIRMOS}. The design of the instrument has been described in prior proceedings\cite{Konidaris2020, Konidaris2022, Konidaris2024}, so here we will provide a broad overview with detailed discussion focused on particular components which have seen significant progress in the last two years. 

\begin{figure}[h]
    \centering
    \includegraphics[width=0.9\textwidth]{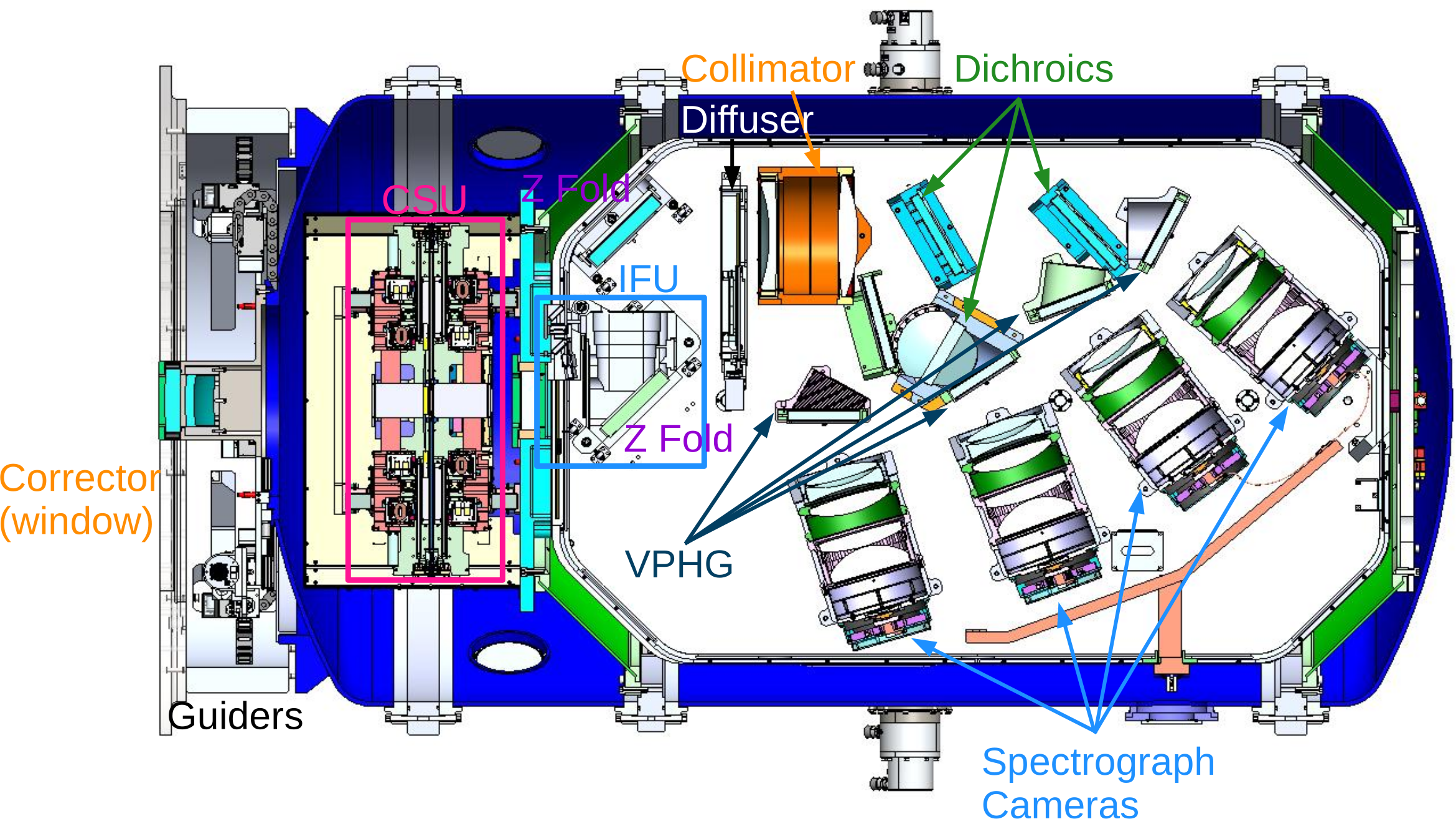}
    \caption{Section view of the MIRMOS instrument model allowing a view of the key components (labeled here). Light enters from the left through a two-element corrector / dewar window. Next, light passes through the slit mask formed by the CSU when operating in long-slit or MOS mode, or the bars open for IFS mode. Next is a field lens embedded in the bench between the CSU and IFU assemblies. In MOS mode (as shown here), the optics of the IFU are moved out of the way by a linear stage. The remaining optics are common, with the slits or pseudo-slits directed by the two Z Fold mirrors into the collimator, before being divided by a dichroic tree to the imaging camera (underneath the optical bench, hidden from view here) or the four VPH gratings and cameras shown here. Each band has a dedicated detector assembly allowing them to operate simultaneously. The exoatmoshpere observations are enabled by inserting an engineered diffuser via a linear stage located just in front of the collimator.}
    \label{fig:MIRMOS}
\end{figure}

A section view of MIRMOS in Figure \ref{fig:MIRMOS} highlights the key components of the design. The multi-object mode utilizes a configurable slit unit (CSU) which can observe up to 92 objects over a $\rm 13'\times3'$ FoV with a mask reconfiguration time of a few minutes or less. Observers can instead choose to utilize the integral field unit (IFU) which sits on a linear stage that will move the optics into the beam for operation with the central CSU bars opened to allow the $\rm\sim18''\times26''$ field to pass through. Both modes share a common collimator and a dichroic tree which splits the light between the single imaging and four spectroscopic channels with independent camera optics, detectors, and VPH gratings. To take advantage of Magellan’s excellent image quality, we will utilize sub-pixel dithering in the spectral and spatial directions. The instrument will achieve its science requirements using its native sampling, but sub-pixel dithering will enable finer sampling at minimal expense by being incorporated in the same mechanism required to provide flexure correction as the instrument rotates.

MIRMOS is designed to observe both faint objects (distant galaxies) with low S/N and very bright objects (stars and their exoplanets) in the high S/N regime. The requirements of observing these two types of objects is, at first glance, orthogonal. For bright objects it is beneficial to oversample the point spread function (PSF) in order to reduce instrument systematics, but for faint objects it is desirable to operate near the Nyquist limit. Both of these requirements can be satisfied in MIRMOS by the inclusion of an engineered diffuser that will be inserted by a mechanical stage for operation in the exoatmosphere mode. This diffuser will broaden the PSF, allowing it to be oversampled while simultaneously mitigating the impact of time-variability due to atmospheric seeing. This also homogenizes the point spread function, making sub-pixel level flat fielding errors negligible. To develop and de-risk this aspect of the design, the \textit{Henrietta} instrument has been developed for the 1m Swope telescope. \textit{Henrietta} is now on-sky undergoing commissioning and is described in detail in prior proceedings\cite{Williams2024}.

\subsection{Optical Design}\label{sec:optics}

\begin{figure}[h]
    \centering
    \includegraphics[width=0.9\linewidth]{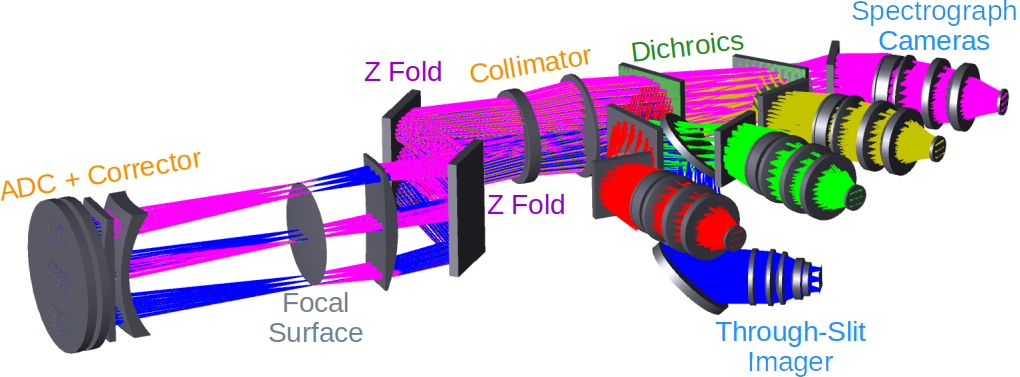}
    \caption{MIRMOS optical layout (with IFU left out for clarity). The ADC prisms are shown on the far left, followed by the corrector/window. The CSU is located at the instrument focal surface and the rest of the optics correspond to those labeled in Figure \ref{fig:MIRMOS}, aside from the through-slit imager (blue rays) which is located on the opposite side of the optical bench from the rest of the components and is therefore hidden from view in the CAD model.}
    \label{fig:optical_layout}
\end{figure}

The MIRMOS optical design utilizes refractive optics aside from fold mirrors (for improved mechanical packaging) and the IFU optics with the layout shown in Figure \ref{fig:optical_layout}. This design work has primarily been carried out using Zemax OpticStudio, with ghost analysis conducted in FRED.

\subsubsection{Performance Budgets}\label{sec:optics_performance}

The MIRMOS image quality requirement for the whole system is to provide $\rm0.36''$ 1D FWHM. Combined with typical seeing at the site, this will produce a $\rm0.4''$ PSF which will be Nyquist sampled by performing subpixel dithering with our $\rm0.4''$ pixels. This performance is achieved through tracking of an error-budget that includes each subsystem contributing to the degradation in performance. This image quality budget is shown in Table \ref{tab:im_qual} and includes an allocation for the contribution of both as-designed and fabrication performance as well as a $\rm10\%$ margin.

\begin{table}[h]
\centering
\caption{MIRMOS Image Quality Budget} 
\label{tab:im_qual} 
\includegraphics[width=\linewidth]{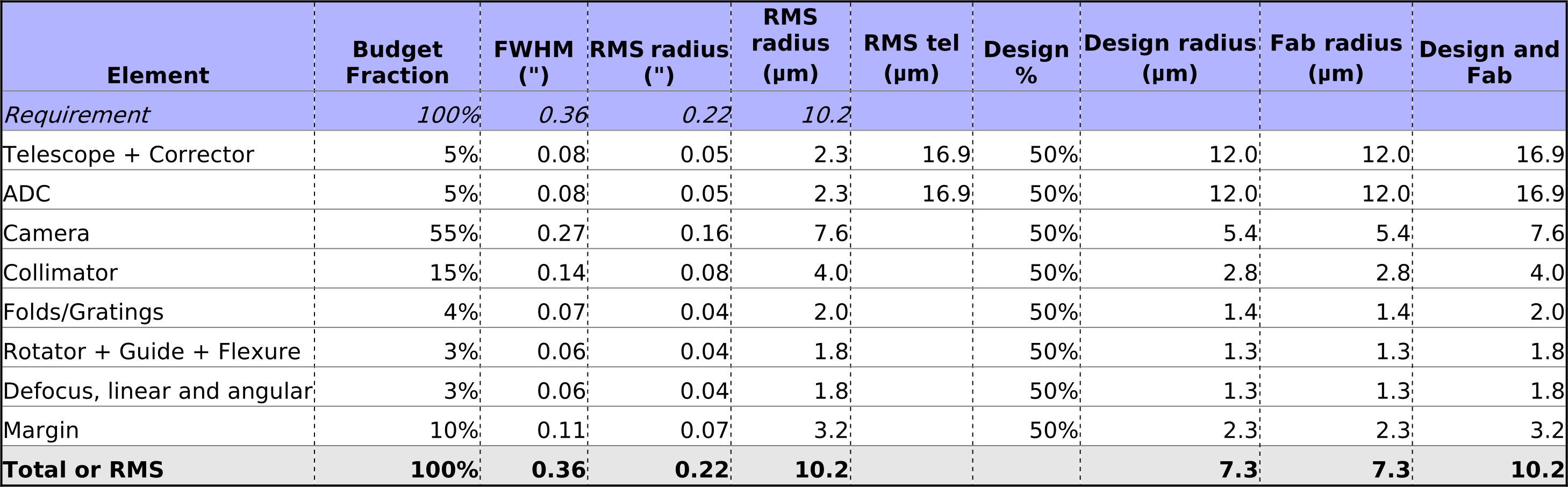}
Note --- Diameters are expressed at the MIRMOS focal plane, except for the lines ``Telescope + Corrector'' and ``ADC'' which are expressed at the telescope focal plane (hence the larger diameters). All subsystems have undergone tolerance analysis and meet the ``Design and Fab" requirement.
\end{table}

The design of these subsystems is described in the following sections. Tolerancing has been performed on these designs with a set of tolerances that encompass both the optical fabrication and optomechanical implementation. Optical fabrication tolerances include material properties (index and dispersion), center thickness, surface curvature, wedge, surface figure error, and, for aspheric surfaces, centering of the surface itself. The optomechanical tolerances encompass assembly and alignment errors like decenter, tilt, and axial location of individual elements and groups of elements.

Initial tolerances were developed using the sensitivity analysis in Zemax and refined iteratively using short-run Monte Carlo analyses. Once a complete set of tolerances is defined, a Monte Carlo analysis with a high run count is executed to develop a statistical sampling of tolerance stack-ups and resulting image performance. As this analysis produces spot sizes that can be asymmetric across the field, we use the average RMS spot radius of 33 locations distributed across the field as the merit function for evaluating each tolerance system. All subsystems have been toleranced individually and meet the specified ``Design and Fab" image quality allocation. A full tolerance analysis of the end-to-end system will be completed.

Like image quality, we track the throughput of each element in the budget shown in Table \ref{tab:throughput}. This budget is very simple, with reasonable estimates for the throughput of each element, since our goal is to maximize throughput balanced with cost.

\begin{table}[h]
\centering
\caption{MIRMOS Throughput Budget} 
\label{tab:throughput} 
\includegraphics[width=.66\linewidth]{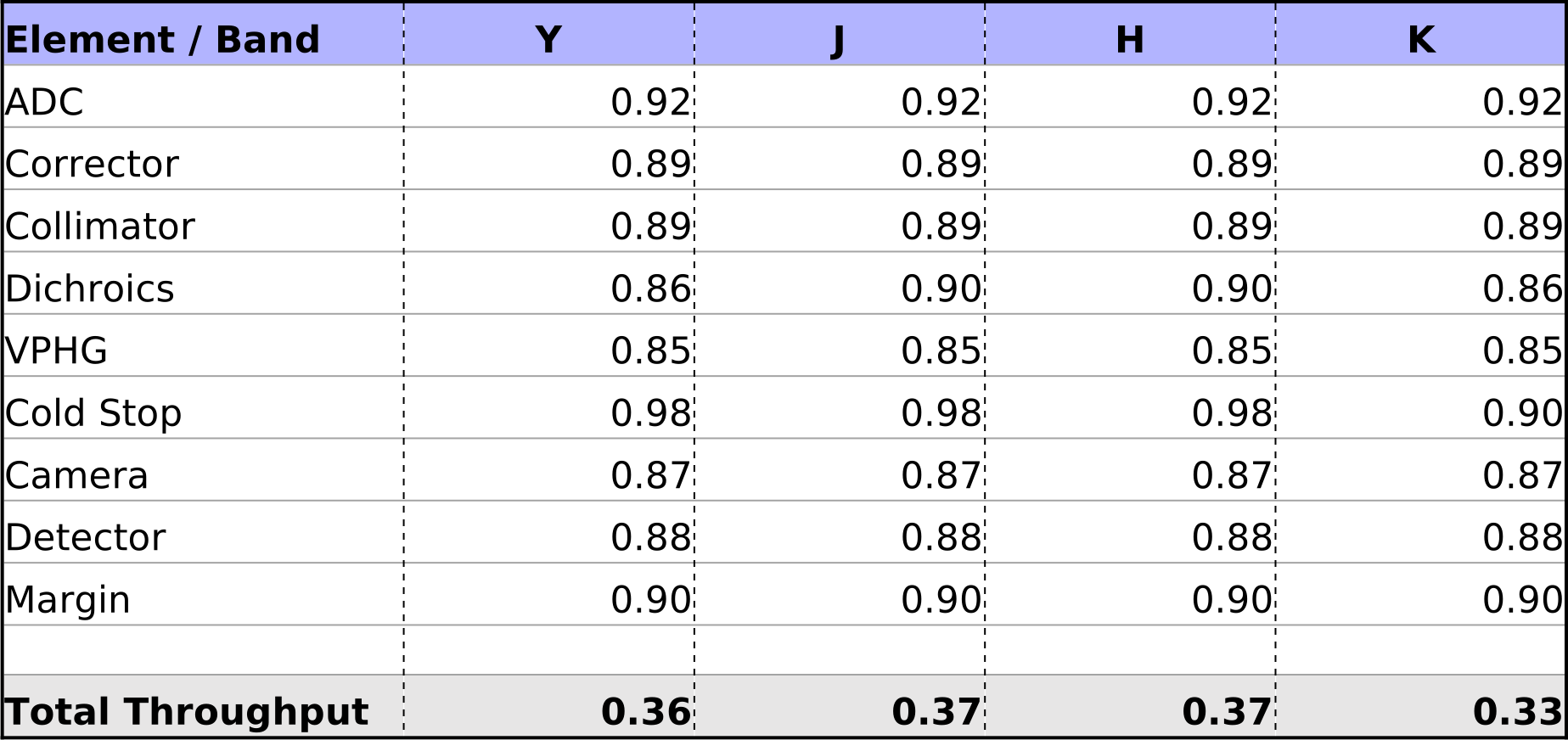}
\end{table}

\subsubsection{Atmospheric Dispersion Corrector}
An Atmospheric Dispersion Corrector (ADC) is required due to our wide simultaneous wavelength range. MIRMOS will utilize a linear ADC consisting of two Fused Silica prisms which piston as the elevation angle of the telescope changes. These optics will sit outside of the instrument dewar due to (i) their size and large range of motion, and (ii) so the guiders see the ADC corrected field.

Since these elements must be warm, background calculations were carried out showing that high quality AR coatings are required to prevent a substantial increase in background at the long end of the K band. Initial discussions have been carried out with vendors, and such 99\% reflectivity coatings for the ADC prisms is feasible at a reasonable cost.

If required -- either for phased delivery or by a particular science case -- MIRMOS can operate with the ADC removed by refocusing the Magellan secondary. This does present an operational challenge, however, since the ADC is mounted inside the elevation disk and is inaccessible without removing MIRMOS. This would therefore be akin to an instrument swap and only potentially be done for prolonged campaigns in an ADC free mode.

\subsubsection{Corrector}
A two element corrector provides sharp images across the FoV of MIRMOS. The front element is also the front window of the vacuum vessel. It will therefore experience stress and deflection due to the pressure difference on either side as well as potentially experience a large temperature gradient. Finite Element Analysis (FEA) was performed on this assembly to simulate the conditions both in lab testing at sea level and in operation at LCO where both the pressure and temperature differential between the inside of the vessel and the outside world will be lower. In both cases the stress was found to be significantly lower than the modulus of rupture for fused silica, with a factor of safety greater than 17. The deformation of the corrector elements has been further modeled with SigFit for incorporation in Zemax to verify that there is no considerable impact on the system performance. This work will be performed during the final design phase.

\subsubsection{Collimator}
The f/11 collimator is a three lens design consisting of a simple fused silica (Suprasil) field lens and an S-FTM16/CaF2 doublet. These materials provide high throughput across the broad wavelength range of MIRMOS. There is a mild asphere on the second surface of the CaF2 lens which has been checked for manufacturability. 

The collimator produces a 118mm beam with 600mm of pupil relief to allow space for the dichroic tree, gratings, and pupil stops. This system has been fully toleranced and meets the subsystem requirement of $\rm4$\um \, RMS radius in the spectral channels (softer images are allowed below $\rm0.886$\um \, in the through-slit imager). A contract is in place with Optimax for the production of the collimator with the material currently in hand. A final update to the design based on precise melt data and index of refraction measurements is discussed further in Section \ref{sec:cryo_indices}.

The expected throughput of the three element collimator has been evaluated based on the glass transmission and the as-designed coating curves. These coatings were designed based on a specification of $\rm R_{avg} < 1\%$ and $\rm R_{max} < 2.5\%$ in the spectral bands (allowed to be looser in the imaging band). The result is an average throughput $\rm >91\%$ in the spectral bands for the range of field angles seen by the collimator, and $\rm >93\%$ in the imaging channel. This exceeds the specified 89\% throughput in the budget of Table \ref{tab:throughput}.

In the K-band, the Lyot stop will be slightly undersized to limit the thermal background. The optimal size for the pupil stop and requirement on pupil centration were determined by simulating the image of the MIRMOS pupil --- including the effects of diffraction --- as well as the thermal background from the telescope+instrument optical path. The S/N is then evaluated for various mask diameters and misalignments, where a misaligned mask leads to accepting thermal emission from outside the pupil (e.g., dome floor, support structure). The best S/N was produced with a mask diameter of 117mm and a misalignment tolerance of $\rm\pm0.1$mm, though this is not a strong function of mask diameter due to a soft pupil image.

\subsubsection{Spectrograph Cameras}
\begin{figure}[h]
    \centering
    \includegraphics[width=0.8\linewidth]{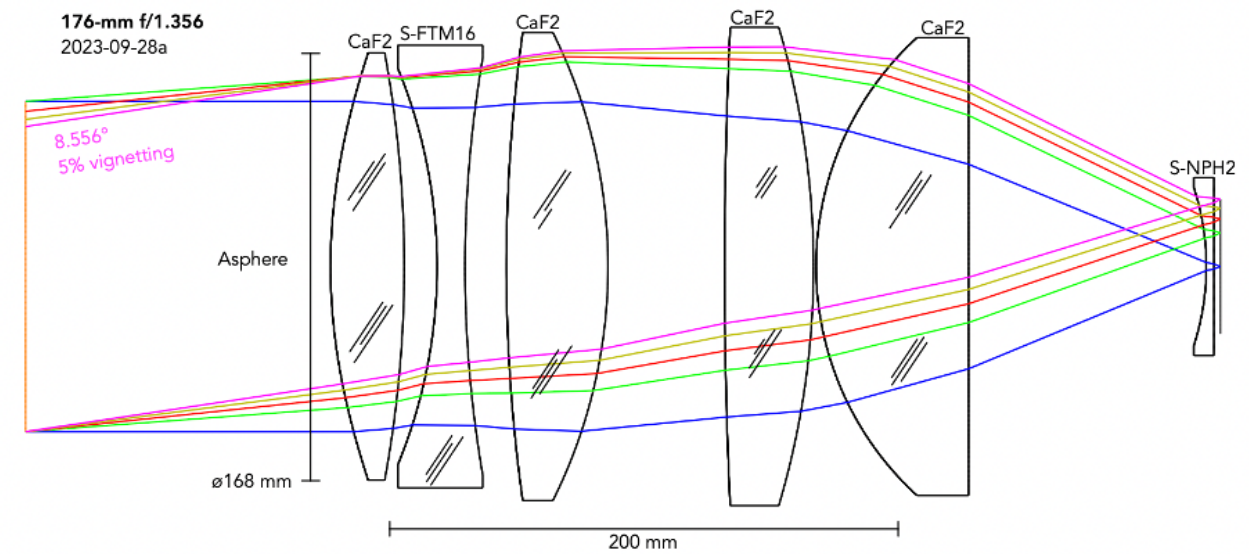}
    \caption{MIRMOS spectrograph camera layout diagram showing only the marginal rays. Ray colors illustrate different field angles with blue corresponding to on axis rays and magenta rays at 8.556$^\circ$ focusing on corner of the detector. A stop at the first surface allows for 5\% vignetting. The second and third lenses have two variations: Y/J and H/K, and the S-NPH2 field flattener varies slightly in all four bands.}
    \label{fig:cam_layout}
\end{figure}

There are four spectroscopic channels in MIRMOS, each requiring an f/1.36 camera with 120mm of pupil relief to limit the amount of 0$\rm^{th}$ order light. Each camera consists of 6 elements with slight variation in the spacing and radii of curvature between bands. The general camera layout is shown in Figure \ref{fig:cam_layout}. The first, fourth, and fifth elements have the same prescription in all bands; elements two and three have a Y/J or H/K variation; and the field flattener (element 6) varies slightly in each band.

These four variations of the camera have undergone a tolerance analysis showing that they will meet the as-built requirement to produce 7.6\um. Since these cameras are not used for imaging, lateral color is allowed with no constraints. Scattered light and ghost analysis have been performed on the Y and H-band cameras as these should be representative of the main differences in the camera variations. This was done in FRED to perform non-sequential raytracing, with baffles matching the preliminary layout and conservative 1.5\% AR coatings on all lens surfaces (the as-designed coatings will have lower reflectivity than this). In this analysis, point source ghosts were attenuated by a factor of $\rm10^{-6}$, and flat field ghosts were at maximum 0.6\% of the average flat field value.

Like the collimator, the optics for each camera have been ordered from Optimax, with a final update of the design based on measured cryogenic index of refraction measurements discussed in Section \ref{sec:cryo_indices}. Based on the coatings designed to meet our specification of $\rm R_{avg}<0.5\%$ (for most surfaces), the cameras are expected to have throughput between $\rm 97 - 91\%$ with the highest throughput in J and the lowest in K-band. For all wavelengths, this is a substantial improvement over the budgeted 87\% for this subsystem.

\subsubsection{Through-slit Imager}
The Through-Slit Imager (TSI) operates from $\rm0.7-0.886$\um \, and is primarily used to verify and maintain slitmask alignment, though it may also be used to study clouds in exoatmospheres while taking transit spectra in the other four channels. 

\begin{SCfigure}[0.8][h]
    \includegraphics[width=0.5\linewidth]{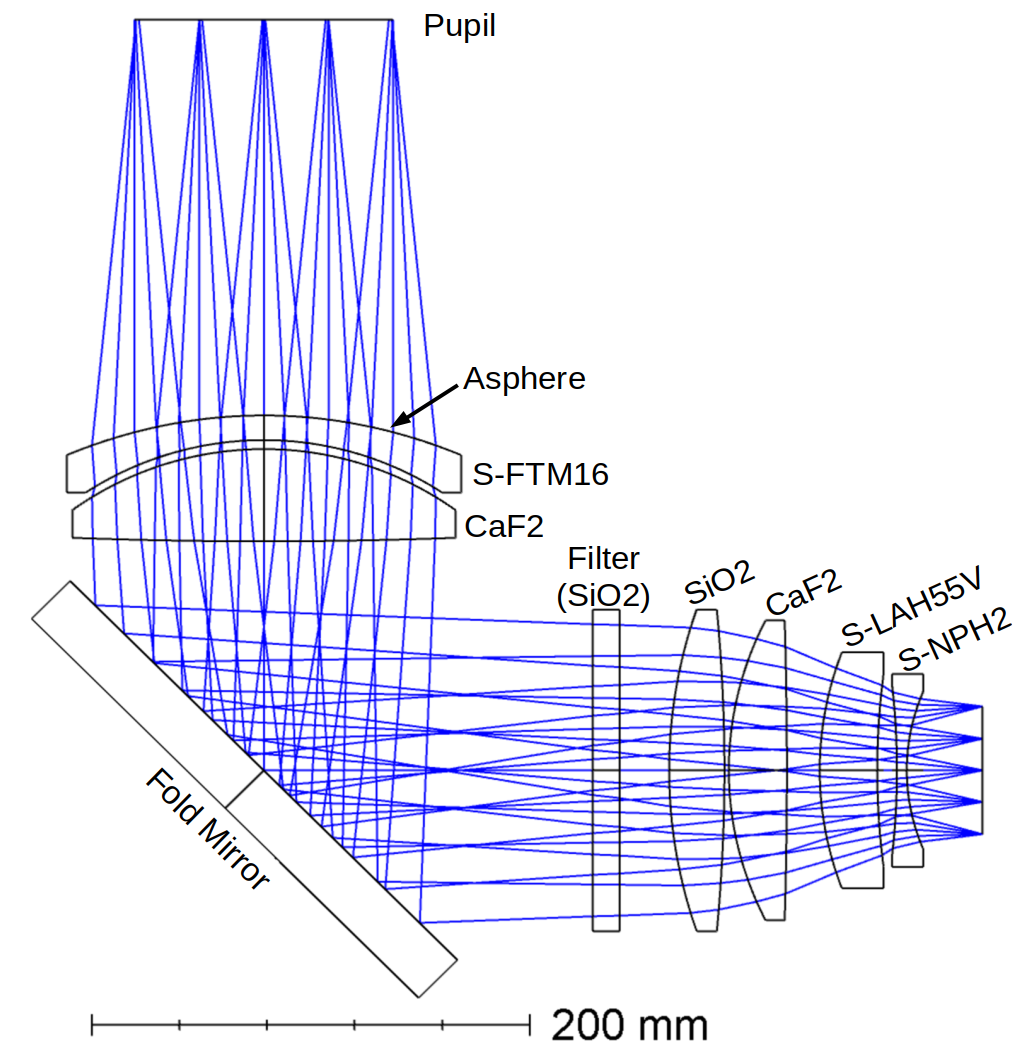}
    \caption{Layout of the MIRMOS Through-Slit Imager (TSI). The camera is split with the first two elements before a fold mirror and the rest after in order to mount the system to the back side of the MIRMOS optical bench. There is an asphere on the first surface of lens 1 with the rest being spherical lenses. There will be a linear mechanism which will select between three possible filter options for target alignment or transit photometry. \vspace{6em}}
    \label{fig:TSI}
\end{SCfigure}

The f/2.5 camera is composed of six lenses packaged with an intermediate fold mirror as shown in Figure \ref{fig:TSI}. These optics are made from S-FTM16, CaF2, Fused Silica, S-LAH55V, and S-NPH2. There is an eighth order asphere on the first surface of lens 1, but no other aspheres in this system. A tolerance analysis has been performed with 90th percentile results for TSI camera exceeding it's ``Design and Fab" image quality requirement of producing $\rm9.8$\um \, RMS spot radii. The TSI will have a filter exchanger with three slots: one broadband and two narrow band filters nominally at the blue and red end of the spectral range (exact specifications are to be determined). We will utilize a frame transfer CCD in order to allow fast readouts for monitoring clouds during an exoplanet transit.

\subsubsection{Integral Field Unit}\label{sec:ifu}

The IFU has undergone a substantial redesign since the last proceedings and a detailed update is provided in this conference\cite{Cosens2026}. In short, the previous system was significantly restricted by mechanical constraints and so the only way to achieve the required image quality was through the use of freeform surfaces\cite{Cosens2024}. As this ended up being too costly to produce we have undergone a full system redesign. The IFU has been allowed to move to a new location in the instrument (shown in Figure \ref{fig:MIRMOS}) which allows the use of a virtual style IFU with no freeform optics. This does require the addition of a re-imaging doublet and two additional reflections, thereby slightly reducing the throughput in this mode, but at the benefit of significantly reducing the cost and simplifying the alignment of the mirror arrays.

\subsubsection{Gratings}
MIRMOS is designed to use Volume Phase Holographic (VPH) gratings. Detailed grating specifications have been produced to facilitate feasibility and performance discussions with potential vendors. Both Wasatch Photonics and INAF's grating laboratory have indicated an ability to meet the provided specifications and provided predicted efficiency curves with average values that are generally above 80\%\footnote{Material absorption above $\sim$2\um \, reduces the predicted efficiency at the red end of K band for both.}. 

Smaller prototype gratings ($\sim$50mm diameter) have already been produced by both of these vendors. The \textit{Henrietta} project is successfully using VPH gratings from Wasatch, and prototype gratings from INAF await cryogenic testing at Carnegie where we will verify that performance is maintained after many thermal cycles.

\subsubsection{Collimator and Camera Final Design} \label{sec:cryo_indices}
\begin{figure}[h]
    \centering
    \includegraphics[width=0.9\linewidth]{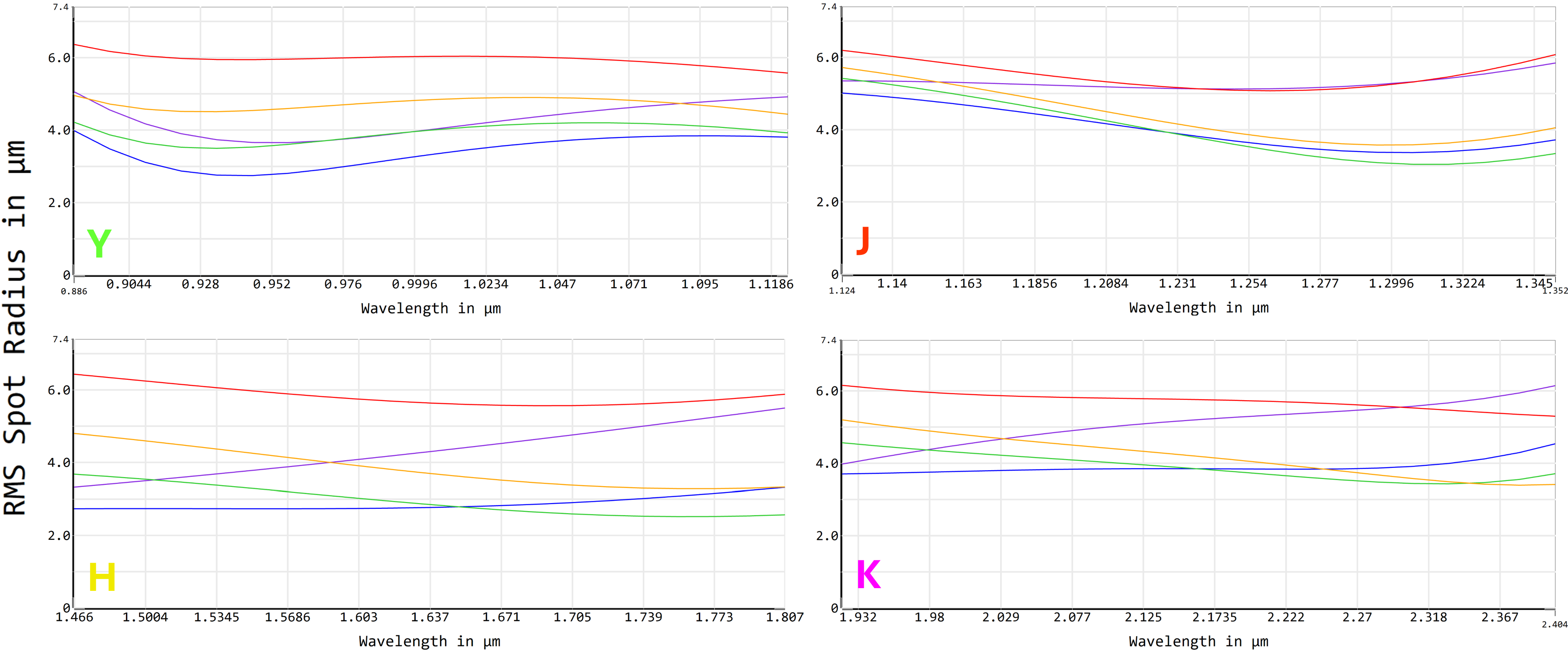}
    \caption{RMS spot radius of the MIRMOS spectroscopic cameras following the measurement of cryogenic indices of refraction. The colored lines correspond to positions from the center to edge of the camera FoV and the color of the band label in the lower left of each plot matches the rays in the layout of Figure \ref{fig:optical_layout}. Shown here is the performance without tolerancing, which does exceed the nominal budget of $\rm5.4$\um \, for this portion of the design across much of the field, however, with tolerances included the ``Design and Fab'' requirement of $\rm7.6$\um \, is being met.}
    \label{fig:cam_performance}
\end{figure}

\begin{SCfigure}[1][h]
    \includegraphics[width=0.6\linewidth]{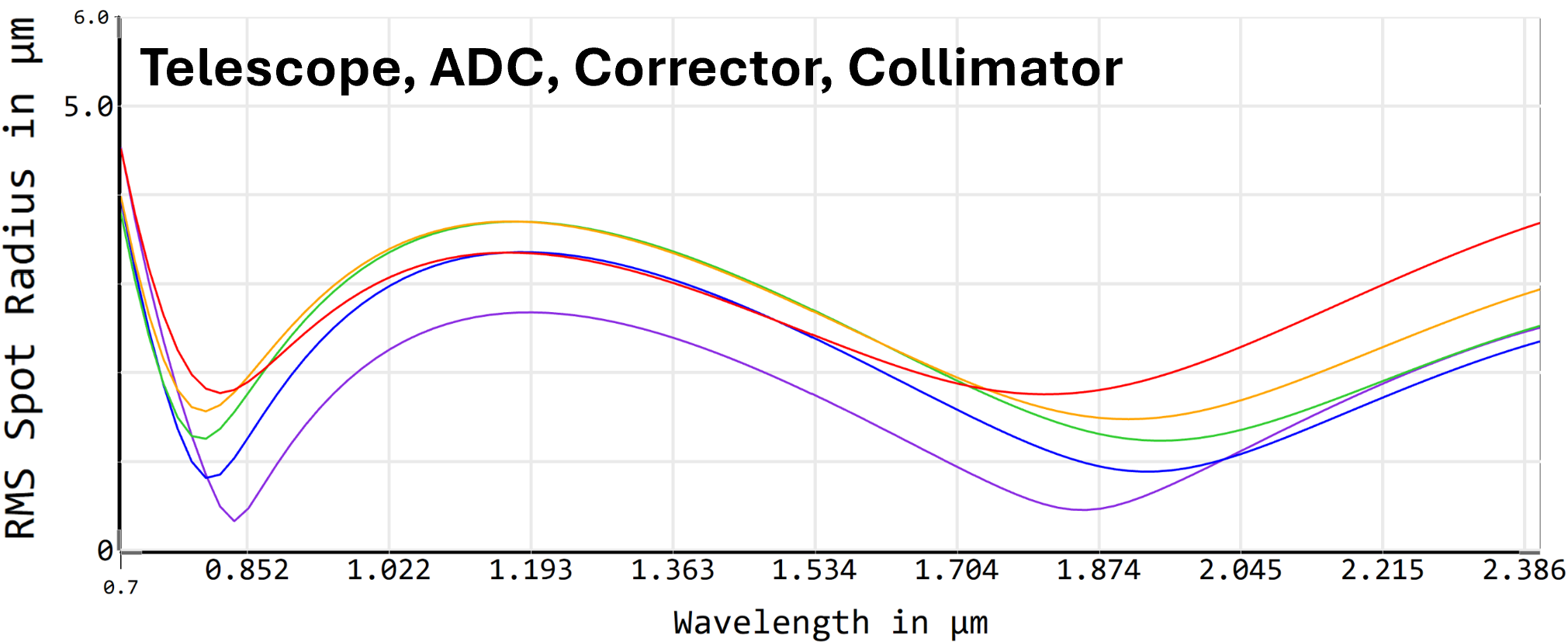}
    \vspace{.8em}
    \caption{RMS spot radius of the MIRMOS system through the collimator optics following the rebalance of the collimator after cryogenic index of refraction measurements. The colored lines correspond to positions from the center to edge of the FoV. Since the collimator was optimized with the telescope, ADC, and corrector optics these spot radii should be compared the the total budgeted impact of these systems ($\rm5.2$\um). This is met across the field and bandpass with the tolerancing of the collimator likewise meeting the ``Design and Fab'' requirement.}
    \label{fig:coll_performance}
\end{SCfigure}

Since the previous SPIE, the optical design of the common elements has been largely completed. Pre-construction designs of the near-IR cameras and collimator lenses were completed and a contract is in place with Optimax for the construction of these optics. This included the production of S-NPH2 and S-FTM16 prisms from the same melt to be used for cryogenic index of refraction measurements at the NASA Goddard Space Flight Center's CHARMS facility. This testing has been completed with the fits provided to the MIRMOS team. Glass catalogs have been created using the measured indices of refraction and incorporated into the MIRMOS collimator and camera optical designs. The designs were then re-optimized. For the cameras, changes were allowed to the lens spacing, radii of curvature, and asphere terms on the first lens element. Constraints were added to maintain adequate edge spacing and focal length, and to not exceed the purchased blank thickness. For the collimator, only curvatures and asphere terms were allowed to vary with a constraint needed on the required blank thickness for the first lens. The performance of these subsystems after the rebalance are shown in Figures \ref{fig:cam_performance} \& \ref{fig:coll_performance}.

These rebalanced designs have also undergone tolerance analysis at the subsystem level. The results of the Monte Carlo runs show that the expected camera and collimator performance will meet the budgeted ``Design and Fab" spot radii in Table \ref{tab:im_qual}.

\subsection{Mechanical Design} \label{sec:mechanical}
The mechanical design of MIRMOS has matured significantly since the last proceedings with the model of the vacuum vessel shown in Figure \ref{fig:MIRMOS}. This vessel mounts to the instrument rotator on the Nasmyth platform and is supported at the rear by a floating support on the MIRMOS cart (Figure \ref{fig:Mag_MIRMOS}). Two guide cameras are mounted on the front of the instrument, each with a two axis stage to control the position of their pickoff mirror.

Inside the cryostat, the large optical bench is supported by two G-10 cones to provide thermal isolation. The bench is held at 120K by two Gifford-McMahon (GM) cryocoolers from Sumitomo. All optical components of MIRMOS are held at 120K, aside from the H2RG detectors which must operate at 80K. These are cooled by a standalone Oxford GM cooler. An additional Oxford GM cooler keeps a zeolite/charcoal getter at 18K and a water trap (also a zeolite/charcoal mixture) at 35K. Multi-layered insulation around the front and rear sections of the optical bench will mitigate the radiative load.

The bench temperature has been simulated in Ansys to verify the cooling time is feasible and check for the presence of temperature gradients. With a bench made of 6061-T6, this results in a 6K gradient from the front to rear of the bench. One option is to instead use 6063-T6 which has similar mechanical properties but improved thermal conductivity at cryogenic temperatures. If this can be sourced at the required size the temperature gradient would be reduced to only 3K. This gradient could also be mitigated with heaters placed around the optical bench. These simulations also showed that a cool down time of approximately one week is achievable with the chosen GM coolers.

\begin{figure}[h]
    \centering
    \gridline{\fig{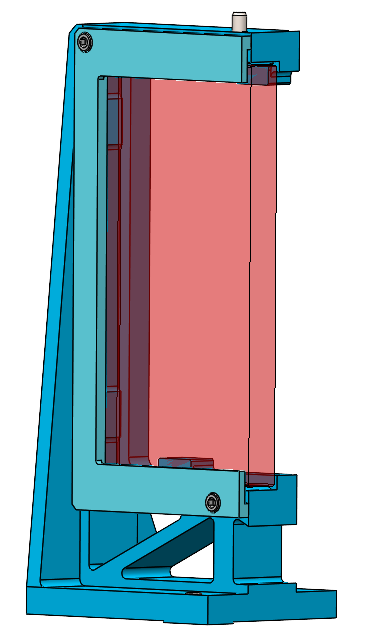}{0.3\textwidth}{}
    \fig{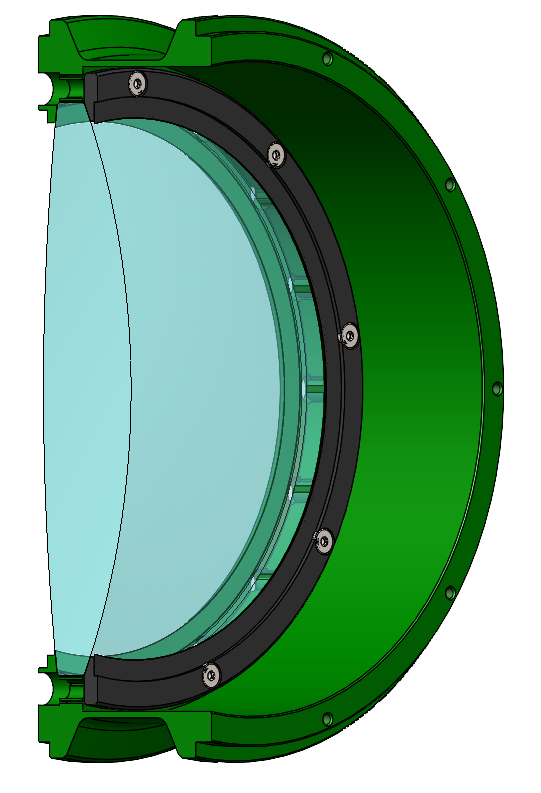}{0.35\textwidth}{}}
    \caption{Section views of MIRMOS optics mounts representative of the two styles used throughout the instrument. On the left is one of the dichroics, representative of the frame style mounts used for the flat optics. The dichroic is held between three pads on the main body of the mount (blue) and corresponding pads on the retainer (lighter blue). On the right is a single specrograph camera lens and its corresponding barrel. Like all the powered optics in MIRMOS (except the IFU), this lens is held radially by a series of roll pin flexures which provide very precise centering. One side of the lens rests on a precisely machined flat surface of the cell (green) which it is held against by a retainer (black) with a pre-load provided by belleville washer stacks.}
    \label{fig:optic_mounts}
\end{figure}

The majority of the MIRMOS optics will be mounted using the roll-pin flexure strategy\cite{Smee2010}. Flat optics with looser centering tolerances (fold mirrors, gratings, and dichroics) will utilize a simpler frame mount. An example of each mounting style is shown in Figure \ref{fig:optic_mounts}. Most of the sub-assemblies shown in Figure \ref{fig:MIRMOS} can be sufficiently located by pins and CNC milling tolerances, but some will require adjustability as outlined in the discussion of alignment in Section \ref{sec:ait}. 

\subsubsection{Cryogenic Mechanisms}\label{sec:mechanisms}

The main optics of MIRMOS are fixed in order to increase stability, but that does still leave the need for some cryogenic mechanisms. These consist of the CSU (being designed by CSEM and described in this conference\cite{Spanoudakis2026}); linear stages for the diffuser, IFU, TSI filter exchanger, and an instrument beam blocker; and a dithering stage for each detector.

\begin{figure}
    \centering
    \includegraphics[width=0.9\linewidth]{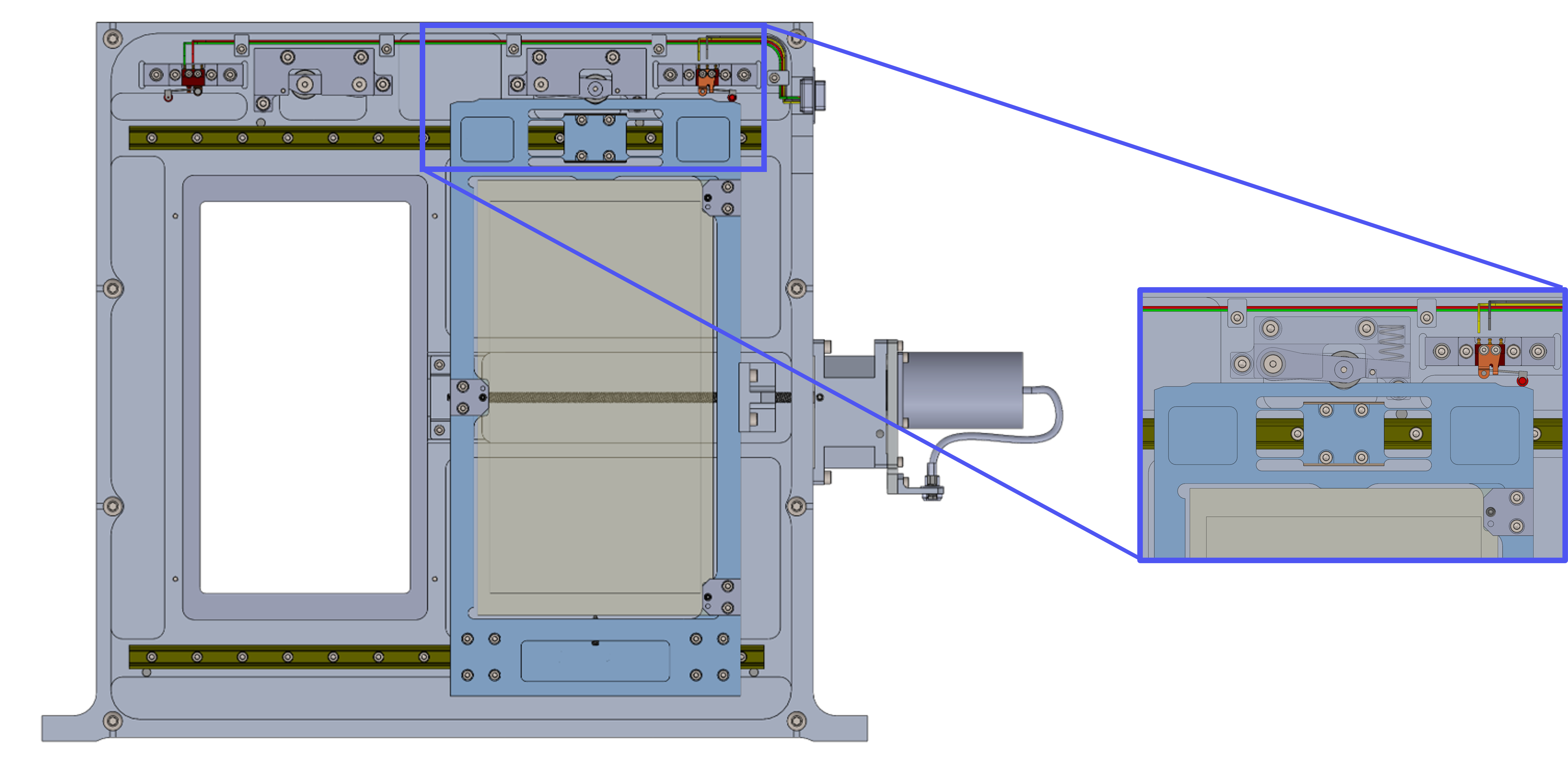}
    \caption{CAD model of the MIRMOS diffuser stage, currently in its stored position. The design shown here is similar across all four linear mechanisms in MIRMOS. It utilizes two parallel PBC Mini Rails with the moving mass fixed via two carriages on the bottom rail and one on the top. A flexure between this carriage and the rest of the diffuser frame prevents the system from binding due to any deviations from parallel or play in the system. The zoom in on the right shows the detent and switch for this position. A second detent and switch are aligned with the active position of the diffuser.}
    \label{fig:lin_mechs}
\end{figure}

The linear stages are designed to take advantage of as much commonality of components and features as possible. They each make use of a Phytron stepper motor, gearbox, and lead screw to move the stage along two parallel PBC rails. Detents engage at each of the nominal positions for improved repeatability and simple switches are used to read the position. An example of one of these stages, which moves the engineered diffuser into the beam or into storage is shown in Figure \ref{fig:lin_mechs}. A prototype of this mechanism is being built now and will undergo lifetime testing when completed (see Section \ref{sec:ait}).

The beam blocker will be nearly identical to this design, just with a simple metal mask in place of the diffuser and frame mount. The TSI filter exchanger will have three possible positions and require longer rails, but otherwise will utilize similar design features. The IFU stage is the most likely to deviate from this baseline design depending on the results of the testing on the diffuser prototype. The PBC linear rails are cost effective, but, as far as we can tell, have not been used in vacuum cryogenic instruments before and so represent somewhat of an unknown quantity. If we find that these are unable to produce the positioning accuracy required over the planned lifetime of the mechanisms, we will instead switch to a more expensive ball guide linear slide from either THK or IKO which have been successfully used in previous projects. This will introduce a CTE difference between the aluminum stage and steel rails which will require a flexure to be implemented to take up this difference. An initial design for such a flexure has already been completed.

Each spectrograph will have an additional mechanism on which the H2RG detector will mount, the Dithering and Flexure Compensation System (DiFCS). This titanium stage will perform high precision x, y motions with a titanium flexure and two piezoelectric actuators. As the name suggests, the DiFCS is designed to correct for flexure in the instrument as well as performing subpixel dithering in order to fully sample the instrument PSF. As such it requires a large range of motion with a high degree of repeatability. This mechanism and the performance of a prototype that has been extensively tested is described in this conference\cite{Smee2026}.

\subsubsection{Flexure and Image Motion Analysis}
There is no de-rotator at the Magellan Nasmyth platforms, and so instead the entire MIRMOS instrument must rotate and will experience a range of gravity vectors during the course of the night. Some flexure is therefore unavoidable, with the DiFCS mechanism at each detector designed to correct the ensuing image motion. To ensure that the range of travel is sufficient to correct the expected flexure, significant effort was put into modeling the MIRMOS bench at various gravity vectors. This type of exercise has been done for previous instruments, but there can be large differences between model predictions and real flexure due to simplifications in the model either due to software limitations and/or computational time required. To generate a high fidelity flexure model for MIRMOS, great care was taken in providing realistic constraints and attachments between modeled parts (an area where significant underestimates of flexure can be introduced) and using sub-models to test the impact of simplifying assumptions.

Finite Element Analysis (FEA) of the full instrument was performed in Ansys with these predicted deflections put into the optical model in Zemax for evaluation of the total image motion. As MIRMOS is a large and complex instrument, a faithful FEA model is computationally expensive. To reduce this complexity and provide validation at intermediate steps the instrument was first broken into sub-models before integration for the final analysis. In the global model, boundary conditions were set to mimic the fixed instrument rotator the front of the vessel will be mounted to and a spring contact for the astatic rear support on the MIRMOS cart. Many of the bolted interfaces at individual components were simplified as bonded, but those critical to the amount of flexure were explicitly modeled: the front and rear bench support interfaces, and the bolted interface of the radiation shield. 

This modeling also showed areas where the stiffness of the optical bench could be improved. One key area of improvement identified was in how the rear of the bench was supported. Originally, the rear support was aligned with the bench itself. Since the vast majority of the instrument components are mounted to one side of the bench, the center of mass was offset from this support. To mitigate the torque this induced on the bench, a stanchion was added to better align the support with the center of mass, providing an 85\% reduction in bench torsion in these simulations. The other main component that was identified to improve the bench stiffness was the radiation shield. This component was redesigned from a simple sheet metal assembly to a bonded panel and frame design which will provide stiffer joints.

\begin{figure}[h]
    \centering
    \includegraphics[width=0.9\linewidth]{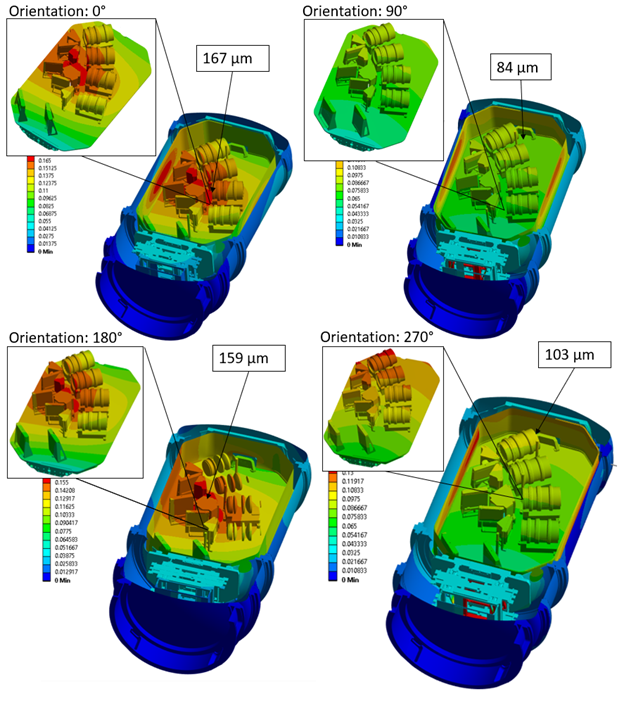}
    \caption{Modeled deflection of the MIRMOS bench at four different gravity vectors. The colorbar is in mm with the maximum deflection called out in each orientation. This model uses the realistic boundary conditions at the instrument rotator and cart as well as realistic bolted interfaces between the bench and support cones, components mounted to the bench, and radiation shield. Results from this simulation are put into Zemax to estimate the amount of image motion that would occur due to these deflections.}
    \label{fig:bench_FEA}
\end{figure}

With these design improvements integrated into the model, the bench deformation was analyzed at four gravity vectors with the realistic interfaces and boundary conditions highlighted above. The FEA results are shown in Figure \ref{fig:bench_FEA}. However, these deflections alone do not linearly translate to the amount of image motion. Instead, the simulations were used to determine the tip, tilt, and focus errors of each optical assembly. These results are then incorporated in the Zemax model of the instrument to evaluate the overall image motion. Initial analysis shows that the expected image motion will be less than half of the required travel range for the DiFCS ($\rm\geq204$\um). As the DiFCS prototype is meeting this requirement\cite{Smee2026}, we expect to have sufficient margin to correct for flexure in the MIRMOS design.

\subsection{Electronics}

\begin{figure}[h]
    \centering
    \includegraphics[width=0.85\linewidth]{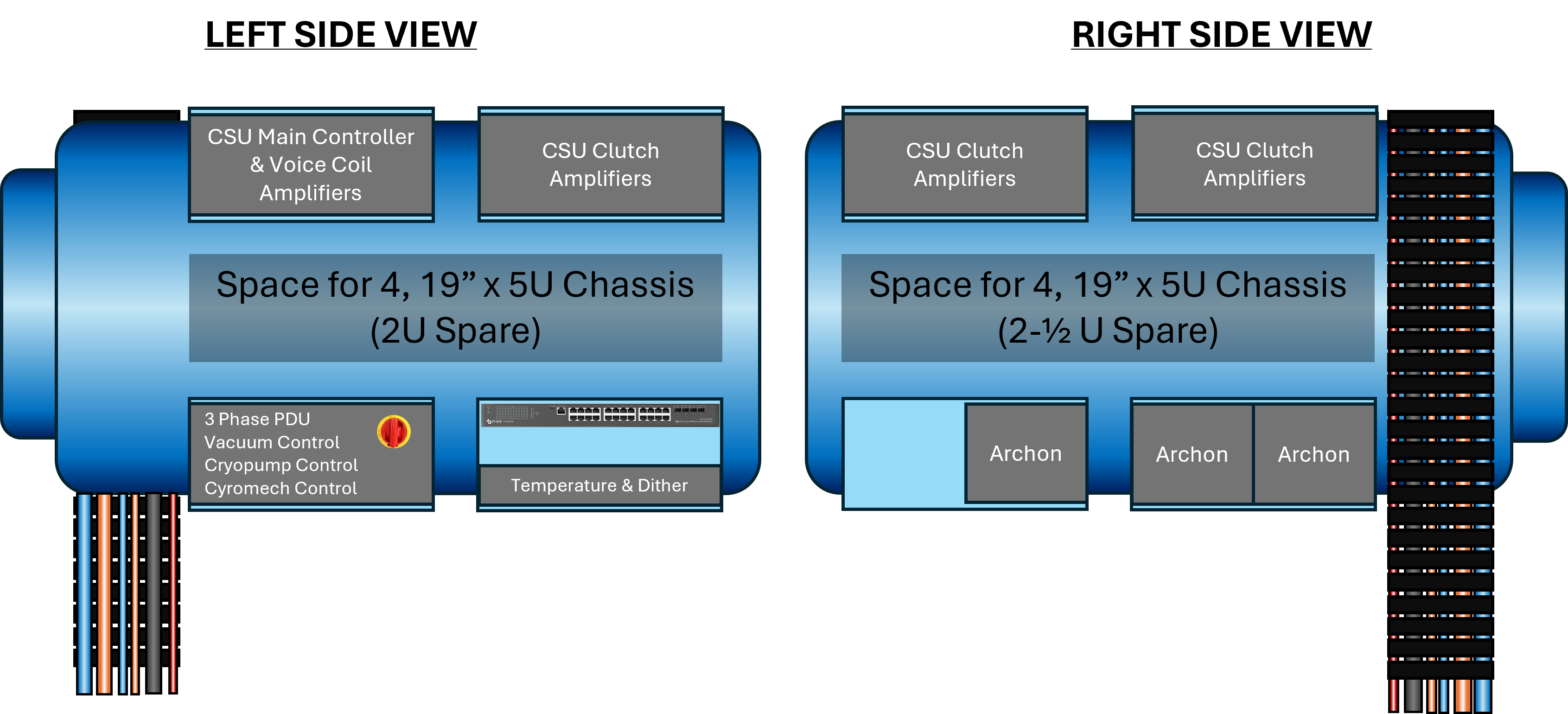}
    \caption{Layout of the on instrument electronics. All of the top racks are occupied by the electronics required to control the CSU. On the bottom we have the electronics required to operate the rest of the instrument including detector controllers (Archon), custom temperature and DiFCS mechanism controllers, and off the shelf controllers for the operation of the GM coolers, vacuum pumps, and other cryogenic mechanisms.}
    \label{fig:electronics_layout}
\end{figure}

Given the limited space on the Nasmyth platform, the MIRMOS electronics have been designed to be compact and make exceptionally efficient use of the space available in the on-instrument racks. The general layout of components within these racks is shown in Figure \ref{fig:electronics_layout}. There are 40 rack units (U's) of space available in these racks, half of which is utilized for the control of the CSU and its 184 individual bars. A cable wrap at the front of the instrument will  contain helium supply and return lines for the GM coolers, glycol lines for cooling the electronics racks, three phase power, and ethernet.

\begin{figure}[h]
    \centering
    \includegraphics[width=0.7\linewidth]{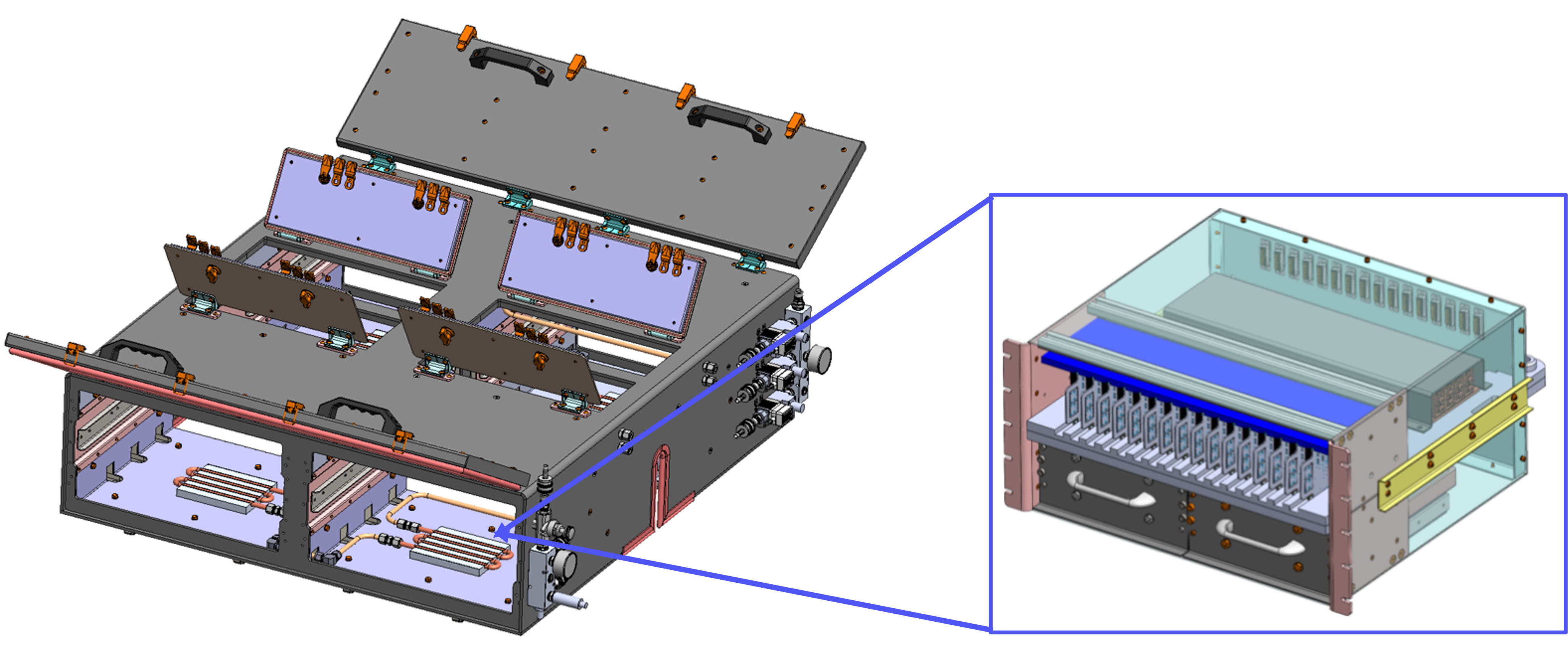}
    \caption{CAD model of the insulated, 20U electronics rack which will mount to the bottom side of the instrument in the cartoon view of Figure \ref{fig:electronics_layout}. There are four, 5U rack spaces (2 front, 2 back) with insulated walls, glycol cooling lines, and access panels. On the left is a model of the rack that will slide into one of these slots. The sheet metal walls are transparent to show the aluminum cold plate at the center with glycol cooling lines built in. This rack will house two battery back-ups on the bottom, and the top will hold the DiFCS and temperature control boards with room for the VCCM behind these.}
    \label{fig:electronics_chassis}
\end{figure}

The ability to keep these electronics on the instrument itself greatly simplifies the cable wrap -- which must accommodate the rotation of the instrument. However, there is requirement to limit heat dissipation into the dome to below 50W so as to not adversely impact seeing. Each rack is therefore thermally insulated with glycol cooling lines throughout the chassis as well as a cold plate which the electronics will be mounted to. An example of one of these racks is shown in Figure \ref{fig:electronics_chassis}.

We plan to control the four GM cryocoolers in MIRMOS (see Section \ref{sec:mechanical}) through the use of Variable Frequency Drives (VFDs) in order to tune (lower) the operating speed. This provides two main benefits. The first is that the lift capacity of the cold heads is selected so as to reduce the time required to reach operating temperature during a cooldown. If these were to continue operating at full speed, a significant amount of heat would need to be added into the system in order to maintain the nominal operating temperature. By instead reducing the operating speed of the cold heads once the nominal temperature is reached, the lift is reduced and less heat needs to be added to into the system. The second benefit is that the wear on the coldheads is reduced, meaning that components do not need to be replaced as often. This is a significant simplification of the maintenance requirements in a system with four GM coolers. Given these benefits, our baseilne plan is to operate the cryocoolers at reduced speeds. The ability to do so with the planned VFDs and coolers will be tested to ensure feasibility.

\begin{figure}[h]
    \centering
    \gridline{\fig{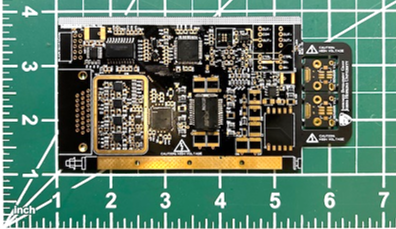}{0.48\textwidth}{DiFCS}
    \fig{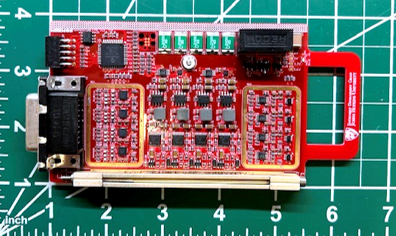}{0.48\textwidth}{Temperature}}
    \caption{Photographs of the prototype DiFCS and temperature control boards. These compact boards control the open loop positioning of both the x and y stages of a single dithering stage (DiFCS boards) or the open loop control of four pairs of heaters and temperature sensors (temp. boards). Sixteen of these boards (including spares) fit in the space allocated to controlling these systems in the rack of Figure \ref{fig:electronics_chassis}.}
    \label{fig:temp_boards}
\end{figure}

The cryogenic mechanisms in MIRMOS will be controlled in two ways for the distinct type of mechanism. We will use Beckhoff PLCs with ethercat modules to control the linear mechanisms and custom controllers for the DiFCS stages. The DiFCS controller has been prototyped along with the mechanical stage and is now on version 2\cite{Smee2024, Smee2026}. They consist of an isolated power supply to drive the piezo stages, an RS485 interface for communication, a pair of 14 bit DAC's for the command position, four 24 bit ADC's for sine-cosine positional feedback, and a microprocessor for PID control. There will be four of these DiFCS controllers, one for each spectroscopic channel. MIRMOS will also utilize custom temperature control boards with much the same architecture as the DiFCS boards. It will also utilize the RS485 communication interface, four PID controllers each taking input from a DAC connected to a four wire temperature sensor and output going to an ADC and low noise voltage controlled heater. These compact DiFCS and temperature control boards are shown in Figure \ref{fig:temp_boards}.

For the four H2RG detectors, we will utilize Archon controllers and custom cryogenic preamplifier boards inside the cryostat which allow the detectors to be read out in buffered mode. This also allows us to use the same controller for the CCD used in the TSI. Since a single Archon can operate two detectors at once, we need a total of three (two for the H2RGs, one for the CCD.). Henrietta served as a testbed for this architecture. The H2RG is driven in 32 channel buffered mode by $\rm100\mu$A current regulating diodes. The buffered outputs are readout differentially by 64 OPA2863A operational amplifiers, chosen for their low drift and low offset voltage/input bias. The differential outputs are then sampled at a rate of 12.5MHz by the Archon’s ADC and many samples are averaged together to create the final pixel value. This readout architecture has proven very successful; achieving correlated double sampling (CDS) read noise of 12$\rm e^-$, low levels of correlated pixel-to-pixel noise that are well corrected by reference pixels, and little to no pixel-to-pixel and channel-to-channel crosstalk. These results will be published in a future manuscript. MIRMOS will adopt a very similar architecture to Henrietta, but with a few improvements: it will use low-voltage differential signaling (LVDS) to provide stable clock signals over large cable distances ($>$1 meter), a shorter distance between the H2RG output and preamp to minimize interference between wavelength channels, and the preamp board will also be cold, providing low noise and high thermal stability. 

\subsection{Assembly and Testing Plan}\label{sec:ait}

Since the last proceedings, we have developed a detailed plan for the assembly and testing of the full instrument and critical subsystems. In preparation for this process we reached out to developers of previous near-IR spectrographs to gather advice on key items to include in our planning. The lessons learned in this process are detailed in this conference\cite{Ramirez2026} and include items we have incorporated in our planning such as the inclusion of more test points to mount temperature probes and the robust testing of all mechanisms as they will be used in the final system.

As part of our testing plan we have procured a large test cryostat which will be used to test the performance and alignment of each camera, measure the wavefront error of the dichroics and gratings, and perform functional and lifetime testing on the cryogenic mechanisms (aside from the CSU which will be tested by CSEM) before incorporation in the full instrument. The test cryostat will include a cart for motorized rotation of the system. The cryostat and cart are being delivered to Carnegie from IRLabs this summer and will first be used to test the prototype diffuser mechanism which uses the same linear rails, stepper motor, switches, and detent design as the other MIRMOS mechanisms. Functional testing at all gravity vectors will be performed as well as full lifetime testing of the mechanism.

\begin{figure}[h]
    \centering
    \includegraphics[width=0.8\linewidth]{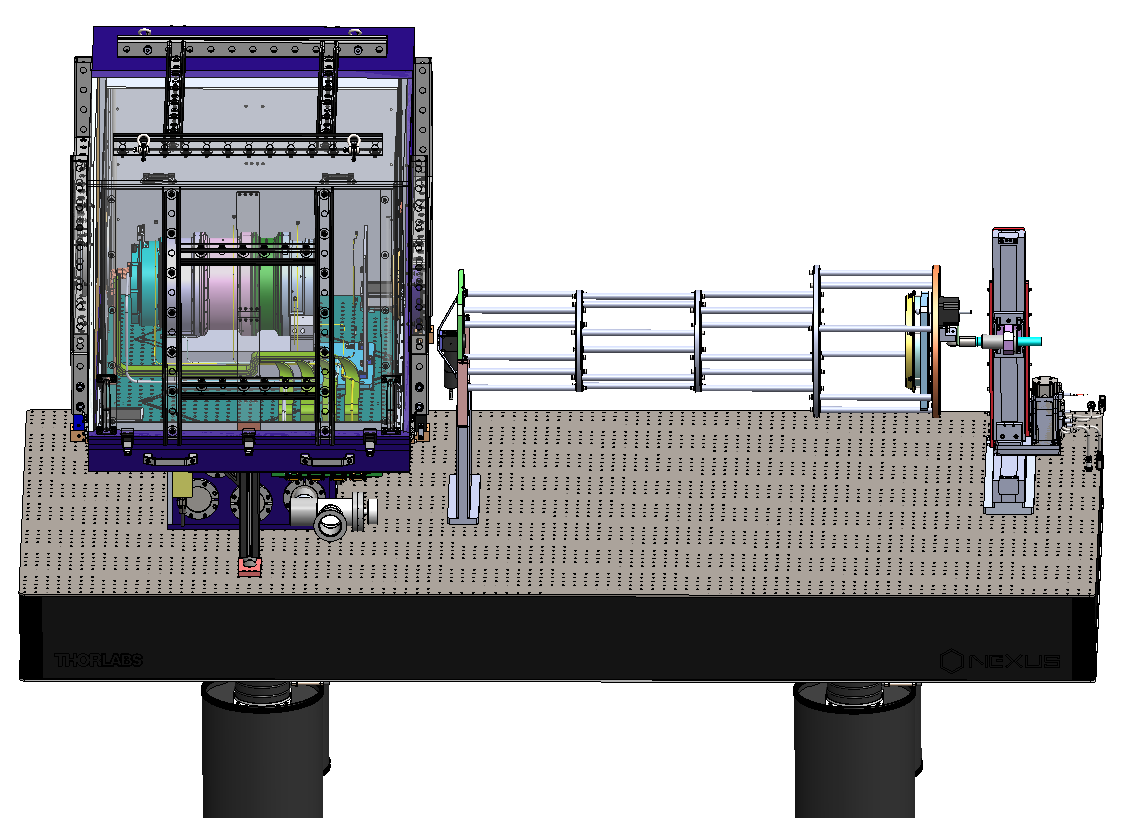}
    \caption{Model of the MIRMOS test cryostat set up for testing of the spectroscopic cameras. The collimator outside the cryostat will be used to measure the focus of each camera as well as image quality at a range of angles to verify the alignment of the complete assemblies. This will be done for all four spectroscopic cameras as well as the TSI.}
    \label{fig:cam_testing}
\end{figure}

Assembly of the spectroscopic cameras will be challenging and is a main risk in MIRMOS. We are using roll-pin flexure style barrels\cite{Smee2010} which we have prototyped in \textit{Henrietta} with lessons learned incorporated into the MIRMOS design. As in \textit{Henrietta}, we will check the centration of each lens element on our Opticentric and gradually build up the cameras. Once assembled, the image quality and focus of each camera will be measured in the test cryostat with a warm collimator aligned on an optical bench (see Figure \ref{fig:cam_testing}). The measured focus will be used to locate the cameras on the main optical bench of the instrument.

As the cameras and mechanisms will be thoroughly tested prior to assembly of the full instrument, this will reduce the number of cooldowns required with the large instrument cryostat. Still, we have allowed time for 11 cooldown cycles in our schedule which includes multiple for contingency. Following the initial qualification of the dewar, we will first align the main optical bench using an alignment telescope and monuments. This will include initial measurement and adjustment warm followed by cold verification. Next the cryogenic mechanisms will be installed and their operation verified. The fold mirrors and dummy optics with cross-hairs will follow to confirm alignment. The fold mirror mounts include adjustability in lateral position as well as tip/tilt for this step. Finally, the real optics, detectors, and all mechanisms will be installed. Not all of these steps need to be sequential meaning that some rearrangement for delivery times is possible without relying on the contingency cooldowns. 

\section{SUMMARY AND PATH FORWARD} \label{sec:summary}
We have described the current design of the Magellan InfraRed Multi-Object Spectrograph (MIRMOS), with a focus on aspects that have seen significant progress since the last update in 2024. MIRMOS will provide spectral coverage from $\rm0.886-2.404$\um \, and imaging from $\rm0.7-0.886$\um \, in either long slit, multi-object, or integral field modes. MIRMOS is currently finishing the preliminary design phase with first-light possible in early 2030 given sufficient funding.

\acknowledgments 
 We thank the entire MIRMOS team and our collaborators at CSEM, Teledyne, and INAF. This material is based on substantial funding from Carnegie Science. This research was funded by the Heising-Simons Foundation through grant 2021-2614. This material is based upon work supported by the National Science Foundation under Grant No. 2206374. We recognize generous support from the Ahmanson Foundation. The Mt. Cuba Astronomical Foundation funded MIRMOS. M.C. is supported by a Brinson Prize Postdoctoral Fellowship.

\bibliography{MIRMOS} 
\bibliographystyle{spiebib} 

\end{document}